\documentclass[a4paper]{article}

\usepackage{INTERSPEECH2020}
\usepackage[export]{adjustbox}
\usepackage{subfigure}
\usepackage{hyperref}
\usepackage[shortlabels]{enumitem}
\hypersetup{
    colorlinks=false,
    linkcolor=blue,
    filecolor=blue,      
    urlcolor=blue,
    citecolor=cyan,
}
\title{Data Efficient Voice Cloning from Noisy Samples with Domain Adversarial Training}
\name{Jian Cong$^1$, Shan Yang$^1$, Lei Xie$^1$$^\dagger$\thanks{$^\dagger$Corresponding author. This research work is supported by the National Key Research and Development Program of China (No.2017YFB1002102).}, Guoqiao Yu$^2$, Guanglu Wan$^2$}
\address{
  $^1$Audio, Speech and Language Processing Group (ASLP@NPU), School of Computer Science, Northwestern Polytechnical University, Xi'an, China \\
  $^2$Meituan-Dianping Group, Beijing, China}
\email{npujcong@mail.nwpu.edu.cn,\{yuguoqiao, wanguanglu\}@meituan.com}

\begin{document}

\maketitle
\begin{abstract}
Data efficient voice cloning aims at synthesizing target speaker's voice with only a few enrollment samples at hand. 
To this end, speaker adaptation and speaker encoding are two typical methods based on base model trained from multiple speakers. The former uses a small set of target speaker data to transfer the multi-speaker model to target speaker's voice through direct model update, while in the latter, only a few seconds of target speaker's audio directly goes through an extra speaker encoding model along with the multi-speaker model to synthesize target speaker's voice without model update. Nevertheless, the two methods need clean target speaker data. However, the samples provided by user may inevitably contain acoustic noise in real applications.  It's still challenging to generating target voice with noisy data.  In this paper, we study the data efficient voice cloning problem from noisy samples under the sequence-to-sequence based TTS paradigm. Specifically, we introduce domain adversarial training (DAT) to speaker adaptation and speaker encoding, which aims to disentangle noise from speech-noise mixture. Experiments show that for both speaker adaptation and encoding, the proposed approaches can consistently synthesize clean speech from noisy speaker samples, apparently outperforming the method adopting state-of-the-art speech enhancement module.
\end{abstract}


\section{Introduction}
Sequence-to-sequence (seq2seq) neural network based text-to-speech (TTS) is able to synthesize natural speech without a complex front-end analyzer and an explicit duration module~\cite{shen2018natural}. However, a sizable amount of high quality audio-text paired data is necessary to build such systems, which limits the model ability to produce natural speech for a target speaker without enough data. Therefore, building target voice with few minutes or even few samples data, or \textit{voice cloning},  has drawn many interests lately~\cite{arik2018neural,yang2016training,cooperzero,jia2018transfer}. In order to produce target speaker voice in a data efficient manner, there are several attempts to build multi-speaker model to produce target voice from a few clean samples, most of which can be divided into two categories~\cite{arik2018neural}: \textit{speaker adaptation} and \textit{speaker encoding}. In both families, a multi-speaker base model is required to generate target voice. 

The core idea for speaker adaptation methods~\cite{chen2018sample,taigman2017voiceloop} is to fine-tune the pre-trained multi-speaker model with a few audio-text pairs for an unseen speaker to produce target voice. The transcription of target speaker samples can be obtained by speech recognition to fine-tune the base model~\cite{huang2020using}. The study in~\cite{moss2020boffin} demonstrates that the training strategy cannot be fixed for adaptation of different speakers and presents a Bayesian optimization method for fine-tuning the TTS model. As for speaker encoding, it mainly builds an extern speaker encoder model to obtain continuous speaker representations for subsequent multi-speaker training. The same extern speaker encoder is then utilized to obtain the speaker embedding from audio samples of an unseen speaker. Without further fune-tuning, the speaker embedding is directly fed into the multi-speaker model to result in target's voice. As the ability and robustness of speaker representation module directly decides the performance of adaptation, several speaker representation methods have been evaluated for adaptive speech synthesis~\cite{cooperzero}. Comparing the above two families, speaker adaptation can achieve better speaker similarity and naturalness, while speaker encoding does not need any extra adaptation procedure and audio-text pairs, achieving so-called one/few-shot(s) voice cloning.

Approaching data efficient voice cloning either through speaker adaptation or via speaker encoding, clean speech samples from target speaker is usually necessary to produce clean target voice. However, in practical voice cloning applications, target speaker data is often either acquired in daily acoustic conditions or found data from Internet, with inevitable background noise. It is still challenging generating target voice with noisy target speaker data, especially for systems built upon the current seq2seq paradigm in which attention-based soft alignment is vunarable to interferences~\cite{yang2020adversarial}. In order to build a robust TTS system, there are several attempts to conduct speech synthesis with noisy data~\cite{yang2020adversarial,valentini2016investigating,hu2019neural}. An alternative method is to de-noise the noisy training data with an external speech enhancement module~\cite{valentini2016investigating}, but the audible or inaudible spectrum distortion may inevitably affect the quality of the generated speech. Besides, we can also try to \textit{disentangle} noise and other attributes in audio. The approach in~\cite{gurunath2019disentangling} aims to disentangle speech and non-speech components using variational auto encoders (VAE)~\cite{doersch2016tutorial} to enable the use of found data for TTS applications. And in~\cite{hsu2019disentangling}, through speaker and noise attributes disentangling during training, the model is able to control different aspects of the synthesized speech with different reference audios. But prior researches on robust TTS have mainly worked on training on large-scale found or noisy dataset, data efficient voice cloning for noisy data has rarely been considered. 
\begin{figure*}[t]
  \centering
  \setlength{\belowcaptionskip}{-0.5cm}
  \setlength{\abovecaptionskip}{3pt}
  \includegraphics[center,width=0.40\paperwidth]{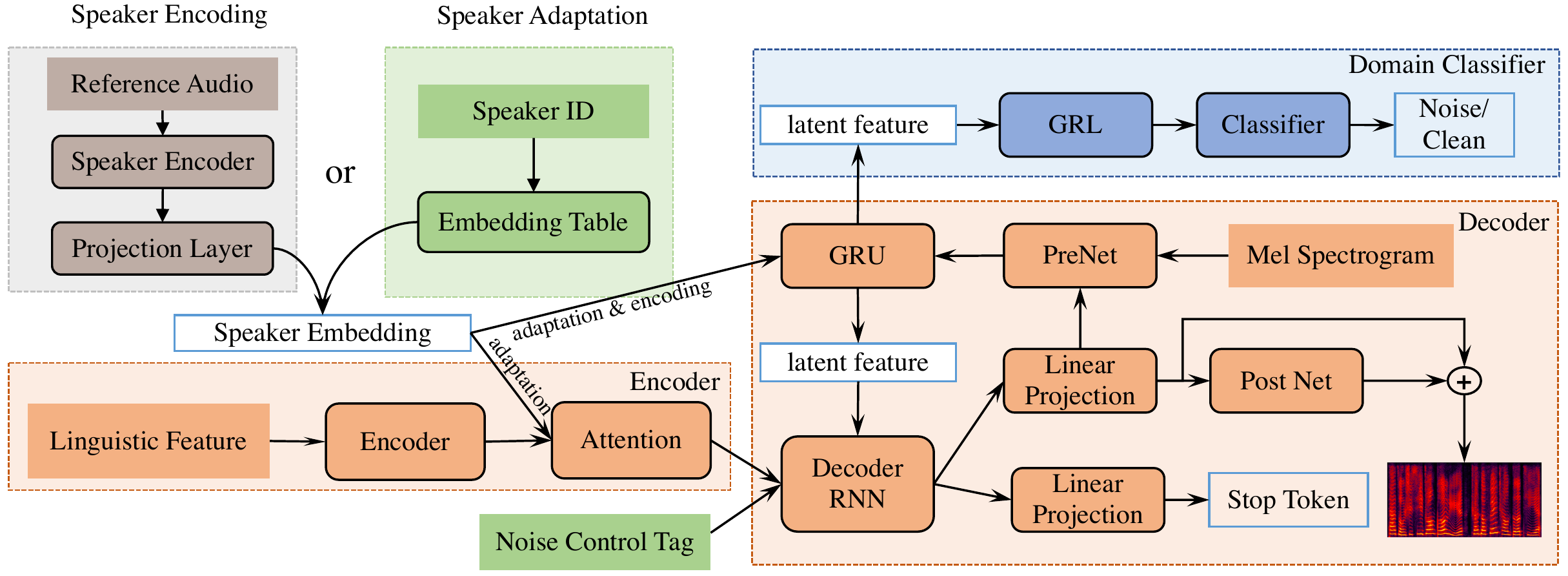}
  \caption{Basic seq2seq TTS model, speaker adaptation and speaker encoding architecture. The components with dotted orange outlines are common basic model with additional GRU. The components with dotted blue outlines are proposed domain classifier module. Speaker adaptation extends basic model with DAT module, speaker embedding looking up table, and noise control tag, shown as green components. Speaker encoding extends basic model with DAT module and external speaker encoding network, shown as gray components. Note that the speaker embedding only injects to GRU at speaker encoding.}
  \label{fig:model_architecture}
\end{figure*}

In this paper, we focus on how to produce target voice in both speaker adaptation and speaker encoding scenes with only a few noisy samples of the target speaker under state-of-the-art seq2seq TTS framework. For the speaker adaptation method, we find the model usually cannot converge at adaptation time with noisy speaker data. So we assume the main challenging problem is how to fine-tune the base multi-speaker model with noisy data and produce clean target speech.  As for the speaker encoding based synthesis model, the main issue is that the speaker representation usually contains noise information, which directly affects the performance of generated speech as the speaker encoding has deviated because of the interference. 

To overcome the above issues in both speaker adaptation and encoding methods, we propose a robust seq2seq framework to conduct target speaker's voice cloning with noisy data. For this purpose, we introduce domain adversarial training (DAT) ~\cite{goodfellow2014generative} to both methods to learn noise-invariant latent features. Specifically, we extend the decoder with a domain classifier network with a gradient reverse layer (GRL) for the speaker adaptation method, trying to disentangle the noise condition in acoustic features. For speaker encoding, since the speaker embedding extracted from the speaker encoder network is noise-dependent, we disentangle the noise condition in the speaker embedding with the help of domain adversarial training, leading to noise-invariant speaker embedding. Note that DAT has been previously studied in speech recognition~\cite{sun2017unsupervised,meng2017unsupervised,sun2018domain}, speaker verification~\cite{wang2018unsupervised,tu2019variational} as well as speech synthesis~\cite{hsu2019disentangling,yang2020adversarial} tasks with superior performance in learning noise-invariant features and attribute disentanglement. To the best of our knowledge, our study in the first one examining its efficacy in data efficient voice cloning. Our study shows that for both speaker adaptation and encoding, the proposed approach can consistently synthesize clean speech from noisy speaker samples, apparently outperforming the method adopting a speech enhancement module.
\section{Proposed Method}
Fig.~\ref{fig:model_architecture} illustrates the proposed seq2seq-based multi-speaker model for data efficient voice cloning in noisy conditions. The proposed architecture contains a CBHG-based text encoder~\cite{wang2017tacotron}, an auto-regressive decoder with GMM-based attention~\cite{battenberg2019location}, the domain adversarial training module, and the speaker representation module. 

For the basic seq2seq framework, the model generates mel-spectrogram $m=(m_1, m_2, \cdots, m_M)$ frame by frame given a text sequence $t=(t_1, t_2, \cdots, t_N)$, where $M$ and $N$ are the length of acoustic features and linguistic features respectively. The text sequence $t$ is firstly fed into the text encoder:
\begin{equation}
  \setlength{\abovedisplayskip}{3pt}
  \setlength{\belowdisplayskip}{3pt}
  x = e(t| \Theta_e)
  \label{eq1}
\end{equation}
where $e(\cdot)$ represents the text encoder and $x$ is the text representation from the encoder.

During the auto-regressive process, the decoder takes current frame of spectrogram $m_t$ to produce next frame $m_{t+1}$. In detail, the decoder firstly converts $m_t$ into latent representation $z_t$ through a pre-net $h(\cdot)$, where the $z_t$ acts as an information bottleneck. The $z_t$ is then treated as a query to compute context vector $c_t$ with $x$ through GMM-based attention module $g(\cdot)$. Hence, the next frame $m_{t+1}$ can be calculated from the context vector $c_t$ and $z_t$ through transformation function $f(\cdot)$:
\begin{equation}
  \setlength{\abovedisplayskip}{3pt}
  \setlength{\belowdisplayskip}{3pt}
	z_{t} = h(m_{t}| \Theta_{h})
  \label{eq21}
\end{equation}
\begin{equation}
  \setlength{\abovedisplayskip}{3pt}
  \setlength{\belowdisplayskip}{3pt}
  c_t = g(z_{t},x| \Theta_{g})
  \label{eq2}
\end{equation}
\begin{equation}
  \setlength{\abovedisplayskip}{3pt}
  \setlength{\belowdisplayskip}{3pt}
  \hat{m}_{t+1} = f(c_t, z_{t} | \Theta_{f})
  \label{eq3}
\end{equation}
where $\Theta_{h}$, $\Theta_{g}$ and $\Theta_{f}$ represent the module parameters of pre-net, attention mechanism and transformation, respectively. We minimize the mean square error between predicted $\hat{m}_t$ and ground truth $m_t$ to optimize the whole model:
\begin{equation}
  \setlength{\abovedisplayskip}{3pt}
  \setlength{\belowdisplayskip}{3pt}
  L_{rcon} = ||m - \hat{m}||_1.
  \label{eq4}
\end{equation}
\subsection{Few-shots robust speaker adaptation with DAT}
To conduct speaker adaptation for noisy data, we firstly build a multi-speaker model with both noisy and clean speech samples. Based on the basic architecture, we adopt an extra trainable speaker embedding table to bring speaker identity. For each speech sample $m$, the speaker representation $s$ is obtained from the embedding table indexed by the corresponding speaker label. We concatenate the speaker embedding $s$ with pre-net and encoder output, so Eq.~(\ref{eq2}) and Eq.~(\ref{eq3}) become
\begin{equation}
  \setlength{\abovedisplayskip}{3pt}
  \setlength{\belowdisplayskip}{3pt}
  z_{t} = h(m_{t}, s| \Theta_{h})
  \label{eq5-1}
\end{equation}
and
\begin{equation}
  \setlength{\abovedisplayskip}{3pt}
  \setlength{\belowdisplayskip}{3pt}
  c_t = g(z_{t},x, s| \Theta_{g}).
  \label{eq5-2}
\end{equation}
In order to build a robust multi-speaker model for few-shots noisy samples in the adaptation stage, we use both clean and noisy speech data during training. As shown in Eq~\eqref{eq5-1}, the latent feature $z$ may contain noise interference when $m$ is noisy. In order to encourage $z$ to become noise-independent feature, we inject a GRU layer into the $h(\cdot)$ and then employ a domain classifier with gradient reversal layers (GRL) on the output $z$ of GRU layer at frame level. The proposed latent $z$ is adopted to predict the noisy/clean label for the following domain classifier. We further feed noisy/clean embedding vector into decoder RNN to control the generation process. With the auxiliary classifier, the final loss function in Eq~\eqref{eq4} becomes:
\begin{equation}
  \setlength{\abovedisplayskip}{3pt}
  \setlength{\belowdisplayskip}{3pt}
  L = L_{rcon} + \lambda L_{noise\_ce}
  \label{eq7}
\end{equation}
where $\lambda$ is the tunable weight for domain classifier loss. 
\begin{figure}[t]
  \centering
  \setlength{\belowcaptionskip}{-1pt}
  \setlength{\abovecaptionskip}{3pt}
  \includegraphics[width=0.9\linewidth]{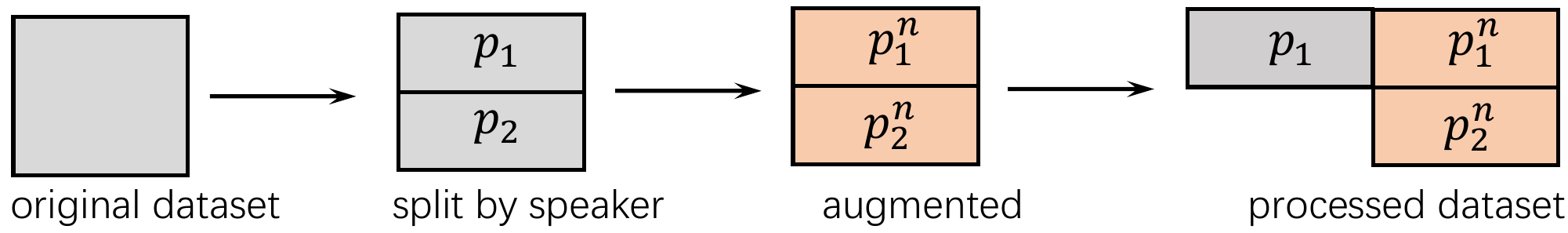}
  \caption{Data augmentation process for speaker adaptation base model.}
  \label{fig:data_aug}
\end{figure}
In order to obtain a multi-speaker corpus for the above domain adversarial training, we apply data augmentation on a clean multi-speaker dataset, as shown in Fig.~\ref{fig:data_aug}. Specifically, we split the original training set into two subsets $p_1$ and $p_2$, both of which contains multiple speakers. Then we add randomly selected background noise at a random signal-to-noise ratio (SNR) in $p_1$ and $p_2$ to obtain the noisy counterparts $p_1^n$ and $p_2^n$. Finally, the subsets $p_1$, $p_1^n$ and $p_2^n$ are treated as the training set to train the above multi-speaker model. Note that there are no clean speech for the speakers in sub-set $p_2^n$, which refers to the speaker adaptation scenario with only noisy speech for each speaker.

For few-shots speaker adaptation scene, there are only several noisy audio clips with transcriptions of the target speaker. We utilize the above noisy samples to fine-tune the pre-trained multi-speaker model with the following steps:
\begin{enumerate}[itemsep=0pt]
  \item Set the noise control tag to `noise' for adaptation data;
  \item Remove the domain classifier loss since we assume the latent $z$ is noise-independent;
  \item Choose a speaker in the training set whose timbre is the most similar to target speaker, and share its speaker embedding to the target speaker~\cite{chen2018sample};
  \item Fine-tune the whole model until convergence;
  \item Set the noise control tag to `clean' and choose the above speaker embedding to generate clean speech of target speaker.
\end{enumerate}
\subsection{One-shot speaker encoding with DAT}
As discussed above, the proposed few-shots speaker adaptation method requires a few adaptation samples with transcription to fine-tune the model. We further propose a robust one-shot speaker adaptation method for noisy target speaker speech without transcription. To this end, we firstly build an individual text-independent speaker discriminative model trained on speaker verification dataset~\cite{cooperzero,jia2018transfer,snyder2018x}. The model adopts time delay neural network (TDNN)~\cite{snyder2018x} to extract the speaker representation (so-called \textit{x-vector}) in the latent space. With the speaker recognition model, we can easily obtain the continuous speaker embedding $s$ for both training and adaptation samples.

Different from the above few-shots speaker adaptation, the noisy target speaker's audio is only used to extract speaker embedding. The domain adversarial training module is the same as previous few-shots adaptation. Since the continuous speaker representation $s$ is obtained from noisy speech, it also inevitably contains noise information. In order to avoid introducing noise into $c_t$, we only inject $s$ in $h(\cdot)$ rather than in both of $g(\cdot)$ and $h(\cdot)$. We train the multi-speaker model with the same objective function in Eq~\eqref{eq7}.

In order to process domain adversarial training, we still need to augment the training set. 
Considering a triple of training samples $<aud_{ref}, text, aud_{tgt}>$, we augment the $aud_{ref}$ with random noise and get $aud^{n}_{ref}$. Therefore, the processed training set is doubled, consisting of two types of samples ($aud_{ref}$ and $aud^{n}_{ref}$) with the same number. And then we apply the same training process as speaker adaptation.

During adaptation, we only need one noisy sample to extract speaker representation $s$ to generate target clean voice. As for a few adaptation samples, we can treat the mean of speaker representations of all sentences as $s$ to control generation, which may be more stable than the $s$ from a single sentence. % Note that we do not need the corresponding transcripts of the speech samples.
\section{Experiments and Results}
\subsection{Basic setups}
\begin{figure}[t]
  \centering
  \includegraphics[width=0.7\linewidth]{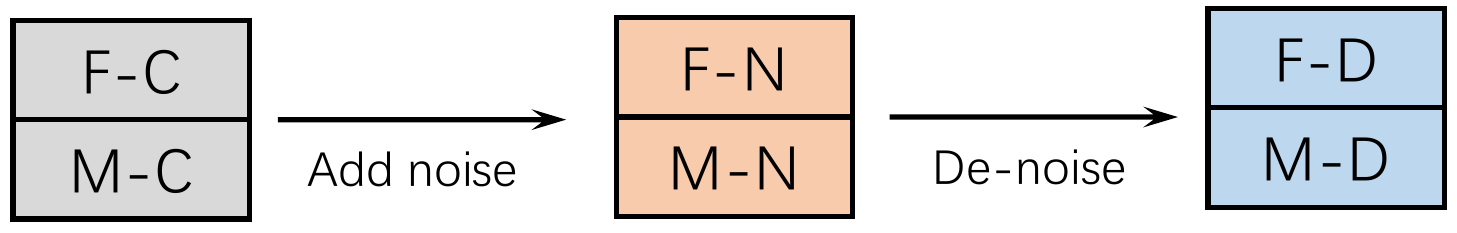}
  \caption{Different parts of test set. F and M indicate female and male, C and N refer to clean and noisy audio, and D indicates de-noised speech.}
  \label{fig:testset}
  \vspace{-0.2cm}
\end{figure}

In our experiments, we use a multi-speaker Mandarin corpus, referred as MULTI-SPK, and a noise corpus from CHiME-4 challenge~\cite{Vincent2017csl} to simulate noisy speech. The MULTI-SPK dataset consists of 100 different speakers in different ages and genders and each speaker has 500 utterances.  The CHiME-4 corpus contains about 8.5 hours of four large categories of background noises. We augment the training set at random signal-to-noise ratio (SNR) ranging from 5 to 25db. We reserve two males (indexed as 001 and 045) and two females (indexed as 077 and 093) as our target unseen speakers for voice cloning experiments. For each target speaker, we select 50 sentences (3-4 minutes of speech) as test samples. The clean test sets for two female and two male speakers are referred as F-C an M-C, respectively. In order to evaluate the performance of noisy target audio, we also add random background noise to F-C and M-C in the way with the training set, resulting in F-N and M-N respectively. As for the de-noising baseline with external speech enhancement module, we use the state-of-the-art speech enhancement model named DCUnet~\cite{choi2019phase} to de-noise F-N and M-N. The internal DCUnet model is trained using over 2000 hours of training data with strong and stable de-noising capacity. The de-noised test sets are referred as F-D and M-D. For clarity, the different parts of test sets are shown in Figure~\ref{fig:testset}. 

To evaluate speaker similarity, we extract x-vectors from the synthesized speech and then measure the cosine distance with the x-vector extracted from original speech of the target speaker. We also evaluate speaker similarity and naturalness using subjective mean score option (MOS) tests, where about 20 listeners examining the testing clips. As for objective evaluation, we measure the mel-cepstral distortion (MCD) between generated and real samples after dynamic time warping.
\subsection{Model details}
All of our models take phoneme-level linguistic features, including phoneme, word boundary, prosody boundary and syllable tone, as input of the CBHG-based encoder~\cite{wang2017tacotron}. The GMM-based monotonic attention mechanism is employed to align phoneme-level linguistic representations and frame-level acoustic features during training~\cite{battenberg2019location}. The architecture of decoder is similar with Tacotron2~\cite{shen2018natural}, and the number of units of additional GRU after pre-net is 256 for latent feature learning. For speaker representation, we adopt straight-forward learnable embedding table for few-shots adaptation, where the dimension of speaker embedding is 256. As for one-shot adaptation, the dimension of x-vector is 512. We concatenate the x-vector with the above latent features.  

For the vocoder, we train gender dependent universal mel-LPCNet, which extends LPCNet~\cite{valin2019lpcnet} with mel-spectrogram, using original MULTI-SPK dataset to convert mel-spectrogram to waveform. The audio samples will be available online\footnote{\url{https://npujcong.github.io/voice_cloning}}.
\subsection{Evaluation on few-shots speaker adaptation}
We firstly train a standard multi-speaker system using original training set without domain adversarial module as our baseline, referred as BASE. As for the robust few-shots speaker adaptation, we propose to conduct adversarial training, referred as DAT.  For the proposed model, we use noise tag (0/1) to control the acoustic condition and the $\lambda$ in loss function is set to 0.1. At adaptation time, we adapt the baseline model BASE and proposed model DAT with different test set, where the batch size is set to 8 and initial learning rate is set to $10^{-5}$. 

Results in terms of various metrics are shown in Table~\ref{table:few-shots} (upper part). For the adaptation with clean target data (F-C, M-C), although we only use half clean speakers in the training set to train the proposed model, the naturalness and similarity of synthesized speech of baseline and proposed model are similar. As for the noisy adaptation data, the BASE model even cannot learn a stable alignment during model fine-tuning, resulting in speech generation failures, i.e. incomplete, mis-pronounced and non-stoping utterances. However, the proposed DAT model still works well to generate target speaker's clean voice, whose performance is close to those samples on clean data in both naturalness and similarity. This result indicates that the proposed approach has ability to produce stable clean target voice under few-shots speaker adaptation scene. We also de-noise the noisy target data to conduct speaker adaptation on the BASE model, but the result indicates that the adaptation with de-noised data suffers from the speech distortion problem, where the MCD is much higher than that of the proposed model. Besides, the similarity is also worse than the proposed model. 
\begin{table}[!htb]
  \centering
  \setlength{\belowcaptionskip}{-4.5pt}
  \caption[]{The results of speaker adaptation and encoding on different test sets and models. “$\times$” means the model is failed to conduct adaptation. N-MOS and S-MOS denote MOS on naturalness and similarity, and SIM-COS is cosine similarity.}
  \resizebox{\linewidth}{!}{
      \begin{tabular}{c|c|c|c|c|c}
      \hline
       \multicolumn{6}{c}{Few-shots Speaker Adaptation}\\\hline
      TEST SET & MODEL & MCD  & N-MOS & SIM-COS & S-MOS \\ \hline
      F-C      & BASE  & 3.77 & 3.42  & 0.96   & 3.64   \\
      F-C      & DAT   & 3.87 & 3.42  & 0.96   & 3.65   \\ \hline
      F-N      & BASE  & $\times$    & $\times$     & $\times$      & $\times$   \\ 
      F-N      & DAT   & 4.18 & 3.36  & 0.95   & 3.65   \\ \hline
      F-D      & BASE  & 4.72 & 3.30  & 0.86   & 3.45   \\ \hline \hline
      M-C      & BASE  & 3.66 & 3.56  & 0.96   & 3.64   \\
      M-C      & DAT   & 3.95 & 3.54  & 0.95   & 3.71   \\ \hline
      M-N      & BASE  & $\times$    & $\times$     & $\times$      & $\times$   \\
      M-N      & DAT   & 4.20 & 3.53  & 0.93   & 3.72   \\ \hline
      M-D      & BASE  & 4.56 & 3.51  & 0.91   & 3.70  \\ \hline\hline
      \multicolumn{6}{c}{One-shot Speaker Encoding}\\\hline
            F-C      & BASE  & 4.30 & 3.39  & 0.91 & 3.51   \\
      F-C      & DAT   & 4.31 & 3.5   & 0.92 & 3.54   \\ \hline
      F-N      & BASE  & $\times$    & $\times$     & $\times$    &$\times$  \\
      F-N      & DAT   & 4.35 & 3.41  & 0.92 & 3.50   \\ \hline
      F-D      & BASE  & 5.32 & 3.32  & 0.89 & 3.4   \\ \hline \hline
      M-C      & BASE  & 4.62 & 3.67  & 0.85 & 3.16   \\
      M-C      & DAT   & 4.58 & 3.63  & 0.88 & 3.26   \\ \hline
      M-N      & BASE  & $\times$    & $\times$     & $\times$    & $\times$  \\
      M-N      & DAT   & 4.55 & 3.67  & 0.88 & 3.34   \\ \hline
      M-D      & BASE  & 4.84 & 3.57  & 0.76 & 2.96   \\ \hline
      \end{tabular}
  }
  \label{table:few-shots}
\end{table}
\subsection{Evaluation on one-shot speaker encoding}
We train an independent x-vector model using internal 3000 hours speaker verification dataset over 2000 speakers. The x-vector is projected to 256 dimension and then used as condition on the TTS model. We train a multi-speaker model with speaker encoder using original dataset as our baseline, referred as BASE and proposed model with DAT using augmented dataset, referred as DAT. During adaptation, we compute mean of x-vectors extracted from 5 sentences randomly selected from test set. Results are shown in the lower half of Table~\ref{table:few-shots}.

For the clean target speakers, we also find there are no significant difference between proposed DAT model and the BASE model. But for noisy data, it is hard to evaluate the performance of the BASE model since it always crashes with the corresponding noisy x-vectors. As for our proposed model, whether the target audio is clean or noisy, it can produce stable and clean synthesized speech of the target speaker. Similar to the speaker adaptation methods for few-shots adaptation, even we de-noise the noisy target audio to extract x-vector, the naturalness and similarity of generated speech is much worse than the proposed DAT method. When we compare speaker adaptation and speaker encoding, we find that speaker adaptation can produce apparently higher speaker similarity samples than speaker encoding. This observation is consistent with~\cite{arik2018neural} as it's still challenging catching speaker's identity in fine-details using just one shot from the speaker; it is even more challenging using one noisy sample.
 
\begin{figure}[t]
  \centering
  \includegraphics[width=0.80\linewidth]{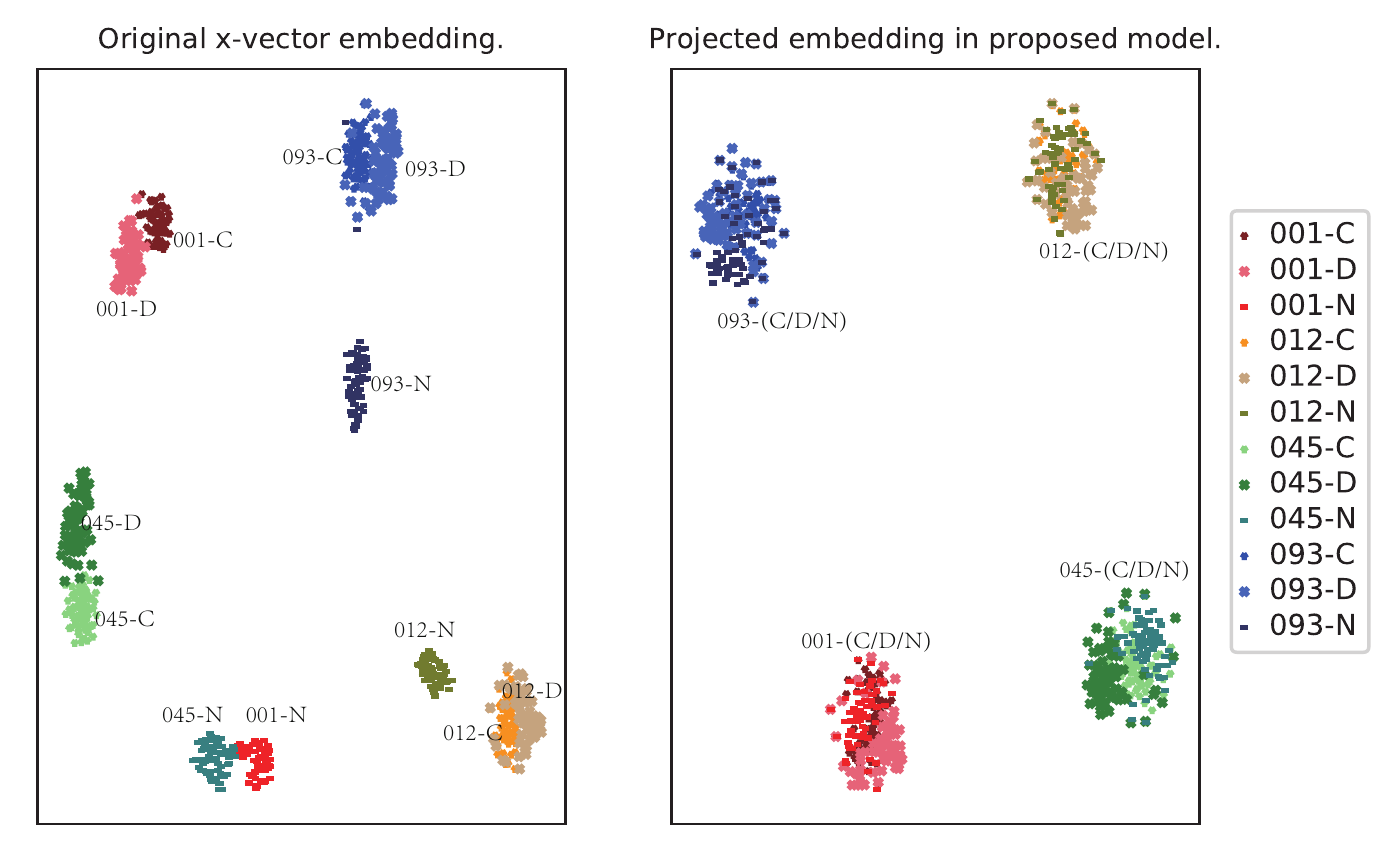}
  \caption{Visualization of utterance embedding from different speakers at clean and noise conditions. C, N and D stand for clean, noisy and de-noised audio.}
  \label{fig:speaker_embedding}
  \vspace{-0.2cm}
\end{figure}

To evaluate the effectiveness of proposed model, we also analyze the projected speaker embedding of our proposed model with the original x-vectors from target speakers using t-SNE~\cite{maaten2008visualizing}, as shown in Fig.~\ref{fig:speaker_embedding}. For x-vectors from target speech, whereas the x-vectors have clear distances between speakers, the speaker representations of noisy and clean samples for the same speaker are usually divided into two clusters. It means that the speaker representation is easily affected by noise interferences, which will directly cause the speaker similarity problem in one-shot speaker adaptation. As for the proposed speaker embedding with adversarial training, we find there is no obvious distance between noisy and clean samples from the same speaker. It indicates that the proposed model successfully disentangles the noise condition from the speaker embedding, which alleviates the negative effects from noise in target speech. 
\section{Conclusions and Future Work}
The paper proposes to use domain adversarial training for data efficient voice cloning from noisy target speaker samples. Results indicate that in both few-shots speaker adaptation and one-shot speaker encoding, the proposed approaches can produce clean target speaker's voice with both reasonable naturalness and similarity. Future work will try to handle the more complicated acoustic condition scenarios, e.g., room reverberations. 
\bibliographystyle{IEEEtran}
\bibliography{mybib}
\end{document}